\documentclass[prb,amsmath,amssymb]{revtex4}
\pdfoutput=1
\usepackage{graphicx}
\usepackage{dcolumn}
\usepackage{bm}

\newcommand\dt
{{\widetilde D}}

\begin{document}

\title{Enhanced reaction kinetics in biological cells}

\author{C. Loverdo}
\author{O. B\'enichou}
\author{M. Moreau}
\author{R. Voituriez}
\affiliation{%
Laboratoire de Physique Th\'eorique de la Mati\`ere Condens\'ee,\\
Universit\'e Pierre et Marie Curie, 4 Place Jussieu, 75252 Paris cedex 05 France }

\maketitle

\textbf{The cell cytoskeleton is a striking example of "active" medium driven out-of-equilibrium by ATP hydrolysis \cite{alberts}. Such activity has been shown recently to have a spectacular impact  on the mechanical and rheological properties of the cellular medium \cite{Thoumine1997,Nedelec1997,Legoff2002,Kruse2004,Storm2005,Voituriez2005,Voituriez2006,Mizuno2007,Dalhaimer2007}, as well as on its transport properties \cite{Sheetz1983,Howard1989,Ajdari1995,Nedelec2001}:  a generic tracer particle freely diffuses as in a standard equilibrium medium, but also intermittently binds  with  random interaction times to motor proteins, which perform active ballistic excursions along cytoskeletal filaments. Here, we propose for the first time an analytical model of transport limited reactions in active media, and show quantitatively how active transport can enhance reactivity for large enough tracers like vesicles. We derive analytically the average interaction time with motor proteins which optimizes the reaction rate, and reveal remarkable universal features of the optimal configuration. We discuss why active transport may be beneficial in various biological examples: cell cytoskeleton, membranes and lamellipodia,  and tubular structures like axons \cite{alberts}.}

 Various motor proteins such as  kinesins or  myosins are able to convert the chemical fuel provided by ATP into mechanical work by interacting with the semiflexible oriented filaments (mainly F--actin and microtubules) of the cytoskeleton   \cite{alberts}.  As many molecules or larger cellular organelles like vesicles, lysosomes or mitochondria, hereafter referred to as tracer particles, can randomly bind and unbind to motors, the overall  transport of a tracer in the cell can be described as alternating phases of standard diffusive transport, and phases of active directed transport powered by motor proteins \cite{alberts,Salman2005,Huet2006}. Active transport in cells has been extensively studied both experimentally, for instance by single particle tracking methods \cite{Sheetz1983,Howard1989}, and theoretically by evaluating the mean displacement of a tracer  \cite{Shlesinger1989,Ajdari1995}, or stationary concentration profiles \cite{Nedelec2001}.

On the other hand, most of cell functions are regulated by coordinated chemical reactions which involve  low concentrations of reactants (such as ribosomes or vesicles carrying targeted proteins), and which are therefore limited by transport. However, up to now
a general quantitative analysis of the impact of active transport on reaction kinetics in cells, and more generally in generic active media, is still missing, even if a few specific examples have been tackled \cite{Holcman2007}. In this letter, we propose an analytical model which allows us to determine for the first time the kinetic constant of transport limited reactions  in active media.

The model relies on the idea of intermittent dynamics introduced  in the context of search processes\cite{nousprotein,SlutskyBiophys04,nousanimaux,Lomholt2005,Benichou2006,Shlesinger2006,Benichou2007,Eliazar2007,Kolesov2007}. We consider a tracer particle evolving in a $d$--dimensional space (in practice $d=1,2,3$) which performs thermal diffusion phases of diffusion coefficient $D$ (denoted phases 1),
 randomly interrupted by  ballistic excursions bound to motors (referred to as phases 2) of
constant velocity $v$ and direction pointing in the solid angle $\omega_{\bf v}$  (see figure 1a). The distribution of the filaments orientation is denoted by $\rho(\omega_{\bf v})$, and will be taken as either  disordered or polarized (see figure 1a), which schematically reproduces the different states of the cytoskeleton \cite{alberts}. The random
duration of each phase $i$ is assumed to be exponentially distributed with mean
$\tau_i$.
The tracer $T$ can react with reactants $R$ (supposed immobile) during  free diffusion phases 1 only, as $T$ is assumed to be inactive when bound to motors, which is realized for instance when reactants are membrane proteins (see figure 1b,c). Reaction occurs with a finite probability per unit of time $k$ when the tracer-reactant distance is smaller than a given reaction radius $a$. In what follows we explicitly determine the kinetic constant $K$ of the reaction $T+R\to R$.

We now present the basic equations  in the case of a   reactant centered in a spherical
domain of radius $b$ with reflecting boundary.  This
geometry both mimics  the relevant situation of a single target and provides a mean field approximation of the general case of randomly located reactants with concentration $c=a^d/b^d$, where $b$ is the typical distance between reactants.  We start from a mean field approximation of  the first order reaction constant\cite{Berg1976}  and write   $K=1/\langle t \rangle$, where the key quantity of our approach is the reaction time $\langle t \rangle$ which is defined as the mean first passage time (MFPT) \cite{Redner2001,nature2007} of the tracer at a reactant position uniformly  averaged over its initial
position. For the active intermittent dynamics defined above, the MFPT of the tracer at a reactant position
satisfies the following backward equation \cite{Redner2001}:
\begin{equation}\label{back1}
\left\{
\begin{array}{l}
\displaystyle D\Delta_{\bf r}t_1+\frac{1}{\tau_1}\int(t_2-t_1)\rho(\omega_{\bf v} )d\omega_{\bf v}-k{\rm I}_a({\bf r})t_1=-1\\
\displaystyle {\bf v}\cdot\nabla_{\bf r}t_2-\frac{1}{\tau_2}(t_2-t_1)=-1\end{array}\right.
\end{equation}
where $t_1$ is the MFPT starting in phase 1 at position ${\bf r}$, and $t_2$ is the MFPT starting in phase 2 at
position ${\bf r}$ with velocity ${\bf v}$. ${\rm I}_a$ is the indicator function of the ball of radius $a$. As these equations (\ref{back1}) are of integro-differential type, standard methods of resolution are not available for a general distribution $\rho$.

However, in the case of a \textit{disordered} distribution of filaments ($\rho(\omega_v)=1/\Omega_d$, where $\Omega_d$ is the solid angle of the d--dimensional sphere), one can use a generalized version of the decoupling approximation introduced in  \cite{Benichou2006} to obtain a very good  approximate solution of equations (\ref{back1}), as described in the Methods section.  We present here  simplified expressions of the resulting kinetic constant by taking alternatively the limit $k\to\infty$, which corresponds to the ideal case of perfect reaction, and the limit  $D\to0$ which allows us to isolate the $k$ dependence.

We first  discuss the  $d=3$ disordered case (see figure 1a), which provides a general description of the actin cytoskeleton of a cell in non polarized conditions, or of a generic in vitro active solution.
An analytical form of the reaction rate $K_{3d}$ is given in the Methods section, and plotted in figure 2a,b. Strikingly, $K_{3d}$ can be maximized (see figure 2a,b) as soon as the reaction radius exceeds a threshold $a_c\simeq D/v$ for the following value of the mean interaction time with motors:
\begin{equation}
\tau_{2,3d}^{\rm opt} =\frac{\sqrt{3}a}{vx_0}\simeq 1.078 \frac{a}{v}
\end{equation}
where  $x_0$ is the solution of $ 2 \tanh(x)-2 x+x \tanh(x)^2=0$.
The $\tau_1$ dependence is very weak, but one can roughly estimate the optimal value by $\tau_{1,3d}^{\rm opt}\simeq 6 D/v^2$.
This gives in turn the maximal reaction rate
\begin{equation}
K^m_{3d} \simeq \frac{c v}{a}\,\frac{\sqrt{3}\left(x_0 -\tanh(x_0)\right)}{ x_0^2  },
\end{equation}
so that the gain with respect to the reaction rate $K^{p}_{3d}$ in a passive medium is
 $G_{3d}=K^m_{3d}/K^{p}_{3d}\simeq Cav/D$ with $C\simeq 0.26$.

Several comments are in order. (i) First,  $\tau_{2,3d}^{\rm opt}$ neither depends on $D$, nor on the reactant concentration. A similar analysis for $k$ finite (in the $D\to 0$ limit) shows that this optimal value does not depend on $k$ either, which proves that the optimal mean interaction time with motors is widely independent of the parameters characterizing the diffusion phase 1. (ii) Second, the value $a_c$ should be discussed. In standard cellular conditions $D$ ranges from $\simeq 10^{-2}\ \mu m^2.s^{-1}$ for vesicles to $\simeq 10\ \mu m^2.s^{-1}$ for small proteins, whereas the  typical velocity of a motor protein is $v\simeq 1\ \mu m.s^{-1}$, value which is widely independent of the size of the cargo  \cite{alberts}. This gives  a critical reaction radius $a_c$ ranging from $\simeq 10\  nm$ for vesicles, which is smaller than any cellular organelle,  to $\simeq 10\  \mu m$ for single molecules, which is comparable to the whole cell dimension. Hence, this shows that in such 3--dimensional disordered case, active transport can optimize reactivity for sufficiently large tracers like vesicles, as motor mediated motion permits a fast relocation to unexplored regions, whereas it is inefficient for standard molecular reaction kinetics, mainly because at the cell scale molecular free diffusion is faster than motor mediated motion. This could help justifying  that many molecular species in cells are transported in vesicles. Interestingly, in standard cellular conditions $\tau_{2,3d}^{\rm opt}$ is of order $0.1\ s$ for a typical reaction radius of order $0.1\ \mu m$. This value is compatible with experimental observations \cite{alberts}, and suggests that cellular transport is close to optimum. (iii) Last, the typical gain for a vesicle of reaction radius $a \gtrsim 0.1 \mu m$ in standard cellular conditions is  $G_{3d}\gtrsim2.5$ (see figure 2a,b) and can reach $G_{3d}\gtrsim 10$ for  faster types of molecular motors like myosins ($v\simeq 4 \mu m.s^{-1}$, see \cite{Sheetz1983,alberts}), independently of the reactant concentration $c$. As we shall see below the gain will be significantly higher in lower dimensional structures such as axons.

We now come to the $d=2$ disordered case (see figure 1c). Striking examples in cells are given by  the cytoplasmic membrane, which is closely coupled to the network of cortical actin filaments, or the lamellipodium of adhering cells \cite{alberts}.  In many cases the orientation of filaments can be assumed to be random. This problem then exactly maps on the search problem studied in \cite{Benichou2006}, where the reaction time was calculated. This permits to show that as for $d=3$, the reaction rate $K_{2d}$ can be  optimized in the regime $D/v\ll a\ll b$. Remarkably, the optimal interaction time $\tau_{2,2d}^{\rm opt}$ takes one and the same value in the two limits $k \to \infty$  and  $D \to 0$ :
\begin{equation}\label{grandv}
\tau_{2,2d}^{\rm opt}\simeq \frac{a}{v}(\ln(1/\sqrt{c})-1/2)^{1/2},
\end{equation}
which indicates that again $\tau_{2,2d}^{\rm opt}$ does not depend on the parameters of the thermal diffusion phase, neither through $D$ nor $k$. In the limit $k \to \infty$  one has $\tau_{1,2d}^{\rm opt} = \frac{D}{2v^2}\frac{\ln^2(1/\sqrt{c})}{2\ln(1/\sqrt{c})-1}$, and the maximal reaction rate can then be obtained :
\begin{equation}
K^m_{2d} \simeq \frac{v\sqrt{c} }{2a\sqrt{\ln(1/\sqrt{c})}}.
\end{equation}
Comparing this expression to the case of passive transport yields a gain  $G_{2d}=K^m_{2d}/K^{p}_{2d}\simeq av\sqrt{\ln(1/\sqrt{c})}/(4D)$. As in the  $d=3$ case, this proves that active transport  enhances reactivity for large enough tracers (with a critical reaction radius $a_c\simeq D/v$ of the same order as in the $d=3$ case) such as vesicles. However, here the gain $G_{2d}$ depends on the reactant concentration $c$, and can be more significant : with the same  values of $D$, $v$ and $a$ as given above for a vesicle in standard cellular conditions, and for low concentrated reactants (like specific membrane receptors) with a typical distance between reactants $b\gtrsim 10 \mu m$,  the typical gain is $G_{2d}\gtrsim 8$, and reaches $10$ for single reactants (like examples of signaling molecules).

The case of \textit{nematic order} of the cytoskeletal filaments, which depicts for instance the situation of a polarized cell \cite{alberts}, can be shown to be equivalent in a first approximation to the   1--dimensional case, which is exactly solvable (see figure 1a,b).
The $d=1$ case is also important on its own in cell biology as many 1--dimensional active structures such as axons, dendrites, or stress fibers  \cite{alberts} are present in living cells. As an illustration, we take the example of an axon,  filled with parallel  microtubules pointing their plus end in a direction  ${\bf e}$. We  consider  a tracer particle interacting with both kinesins ("+" end directed motors, of average velocity  $v {\bf e}$ ) and dyneins ("-" end directed motors,  of average velocity $-v {\bf e}$) with the same characteristic interaction time $\tau_2$  (see figure 1b). For this type of tracer, the MFPT satisfies equations (\ref{back1}) with an effective nematic distribution of filaments  $\rho(\omega_{\bf v})=\frac{1}{2}(\delta({\bf v}-{\bf e})+\delta({\bf v}+{\bf e}))$.
The reaction rate $K_{1d}$ is obtained exactly in this  case (see Methods), and is maximized in the regime $D/v\ll a\ll b$  for the following values of the characteristic times (see figure 2c,d)
\begin{equation}\label{opt1d}
\tau_{1,1d}^{\rm opt}=\frac{1}{48} \frac{D}{v^2c}, \;\tau_{2,1d}^{\rm opt}=\frac{1}{\sqrt{3}}\frac{a}{v c^{1/2}},
\end{equation}
for $k\to \infty$. The maximal reaction rate $K_{1d}^{m}$ is then given by
\begin{equation}
 K_{1d}^{m}\simeq \frac{\sqrt{3}vc^{3/2}}{2a},
\end{equation}
 and  the gain  is $G_{1d}=K_{1d}^{m}/K^{p}_{1d}\simeq av/(2\sqrt{3}Dc^{1/2})$, which  proves that active transport can optimize reactivity as in higher dimensions. Very interestingly the $c$ dependence of the gain is much more important than for $d=2,3$, which shows that the efficiency  of active transport is strongly enhanced in 1-dimensional or nematic structures at low concentration. Indeed, with the same values of $D$, $v$ and $a$  as given above for a vesicle in standard cellular conditions, and for a typical distance between reactants $b\gtrsim 100 \mu m$ (like low concentrated axonal receptors), one obtains a typical gain $G_{1d}\gtrsim 100$ (see figure 2c,d). In the limit of finite reactivity ($k$ finite and $D\to 0$) one has $\tau_{1,1d}^{\rm opt} = \sqrt{a/vk} (1/12c )^{1/4}$ and the same optimal value (\ref{opt1d}) of $\tau_{2,1d}^{\rm opt}$. As in higher dimensions  $\tau_{2,1d}^{\rm opt}$  depends neither on the thermal diffusion coefficient $D$ of phases 1, nor on the association constant $k$, which shows that the optimal interaction time with motors  $\tau_2^{\rm opt}$  presents remarkable universal features. Furthermore, our approach permits an estimate of $\tau_2^{\rm opt}$  compatible with observations in standard cellular conditions, which suggests that cellular transport could be close to optimum.

$\ $

\textbf{ Methods}

The approximation scheme to solve the integro-differential equations (\ref{back1}) relies on the  auxiliary function
\begin{equation}
s({\bf r})=\frac{1}{\Omega_d}\int\!t_2d\omega_{\bf v},
\end{equation}
and on the following decoupling hypothesis:
\begin{equation}\label{dec}
\langle v_i v_j t_2\rangle_{\omega_{\bf v}}\simeq\langle v_i v_j\rangle_{\omega_{\bf v}}\langle t_2\rangle_{\omega_{\bf v}}=\frac{v^2}{d}\delta_{ij}s({\bf r}).
\end{equation}
Similar arguments as provided in \cite{Benichou2006} then leads to the diffusion-like equation
\begin{equation}\label{dtilde}
{\widetilde D}\Delta s({\bf r})-\frac{1}{\tau_2}(s({\bf r})-t_1)=-1
\end{equation}
where ${\widetilde D}=v^2\tau_2/d$.
After rewriting equation(\ref{back1}) as
\begin{equation}\label{Dt1}
D\Delta t_1+\frac{1}{\tau_1}(s({\bf r})-t_1)-k{\rm I}_a({\bf r})t_1=-1,
\end{equation}
equations(\ref{dtilde}) and (\ref{Dt1}) provide a closed system of linear differential equations for the
variables $s$ and $t_1$, whose resolution is tedious but standard. For $d=3$, we obtain in the limit of perfect reaction $k\to\infty$ and  low density  $a\ll b$:
\begin{equation}\label{k3d}
K_{3d}\simeq\frac{3\alpha_1\dt\,(D\tau_1T(1+\alpha_1a)+\dt(\tau_1+\tau_2)(\alpha_2a-T))}{b^3(T+\alpha_1\alpha_2\dt(\tau_1+\tau_2))}
\end{equation}
where $\alpha_1=1/\sqrt{D\tau_1},\alpha_2=3/(v\tau_2)$ and $T=\tanh(\alpha_2a)$. This decoupling assumption  has been controlled numerically for a wide range of  the parameters for $d=2$ \cite{Benichou2006}, and  is shown here to  be satisfactory also for $d=3$ (see figure 2a,b).
For $d=1$ the decoupling approximation is exact  and yields after straightforward calculations an explicit though hardly handleable form of the reaction time. In the regime $D/v\ll a\ll b$, we obtain in the limit $k\to \infty$ the simple form
\begin{equation}\label{tsimp}
K_{1d}= \frac{a}{b}\, \frac {6vs\left( a+\sqrt {
{ D\tau_1}} \right)}{\left( 6sv\tau_1 +\sqrt{3ab}\right) \left( a\,(1+4s^2)+4s^2\,
\sqrt {{ D\tau_1}} \right) }
\end{equation}
where $s=\frac{\sqrt{ab}}{v\tau_2\sqrt{3}}$. Optimization is then performed using standard methods of functional analysis.

$\ $

\textbf{ Author information}

 Correspondence and requests should be addressed to R.V. (voiturie@lptmc.jussieu.fr).
 All authors contributed equally to this work.


$\ $

\textbf{Figure legends:}

FIG. 1: Model of reaction kinetics in active media, and examples of low dimensional structures in biological cells. \textbf{a}, The reactant alternates thermal diffusion phases (regime 1 in red)
of mean duration $\tau_1$ and diffusion coefficient $D$, and ballistic phases of velocity ${\bf v}$  powered by molecular motors (regime 2 in blue) of mean duration $\tau_2$. The cytosqueletal filaments (in black) are here in a disordered state. The polarized nematic state would correspond to parallel filaments, and is equivalent in a first approximation  to a 1--dimensional situation  (see \textbf{b}) with the same concentration $c_{1d}=c_{3d}=a_{3d}^3/b_{3d}^3$ and an effective reaction radius $a_{1d}=a_{3d}c_{3d}^{2/3}$. Molecular motors are not represented. \textbf{b}, Tubular structures in cells such as axons and dendrites ($d=1$). \textbf{c}, Planar structures such as  membranes and  lamellipodia ($d=2$).

FIG. 2: Optimization of the reaction rate. \textbf{a,b}, Gain of reactivity due to active transport $G_{3d}$ for $d=3$  as a function of  $\tau_2$ for different values of the ratio $b/a$ (logarithmic scale). The analytical form obtained in the Methods section (plain lines) is plotted against numerical simulations (symbols) for the following values of the parameters (arbitrary units): $a=1.5$ (brown), $a=4.5$ (red), $a=7.5$ (blue), $a=10.5$ (green), $a=15$ (yellow), with $\tau_1=6$, $v=1$, $D=1$. $K_{3d}$ presents a maximum only for $a>a_c\simeq 4$. Standard cellular conditions (as discussed in the text) correspond to green and yellow curves for $b/a=40$.  \textbf{c,d},  Gain of reactivity due to active transport $G_{1d}$ for $d=1$  as a function of  $\tau_2$ (\textbf{c}) and $\tau_1$  (\textbf{d}) (logarithmic scale). The analytical form obtained in the Methods section (plain lines) is plotted against the exact solution (symbols), for the following values of the parameters (arbitrary units): $D=1,v=1$ for all curves and $a=10,b=10^4$ (red), $a=10,b=10^3$ (blue), $a=2.5,b=10^3$ (green). Standard cellular conditions (as discussed in the text) correspond to blue and red curves.

\newpage

\begin{figure}[ht]
\centering\includegraphics[width =1\linewidth,clip]{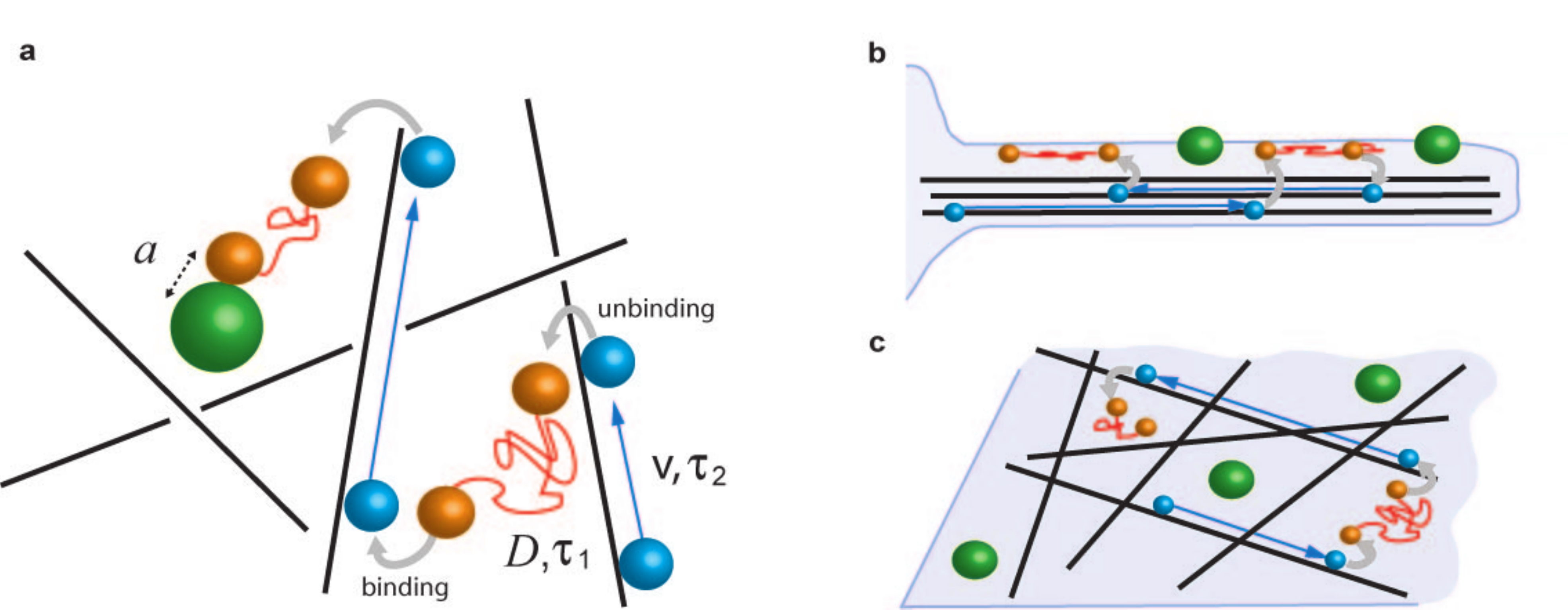}
\caption{}
\label{random05b}
\end{figure}

\begin{figure}[ht]
\centering\includegraphics[width =1\linewidth,clip]{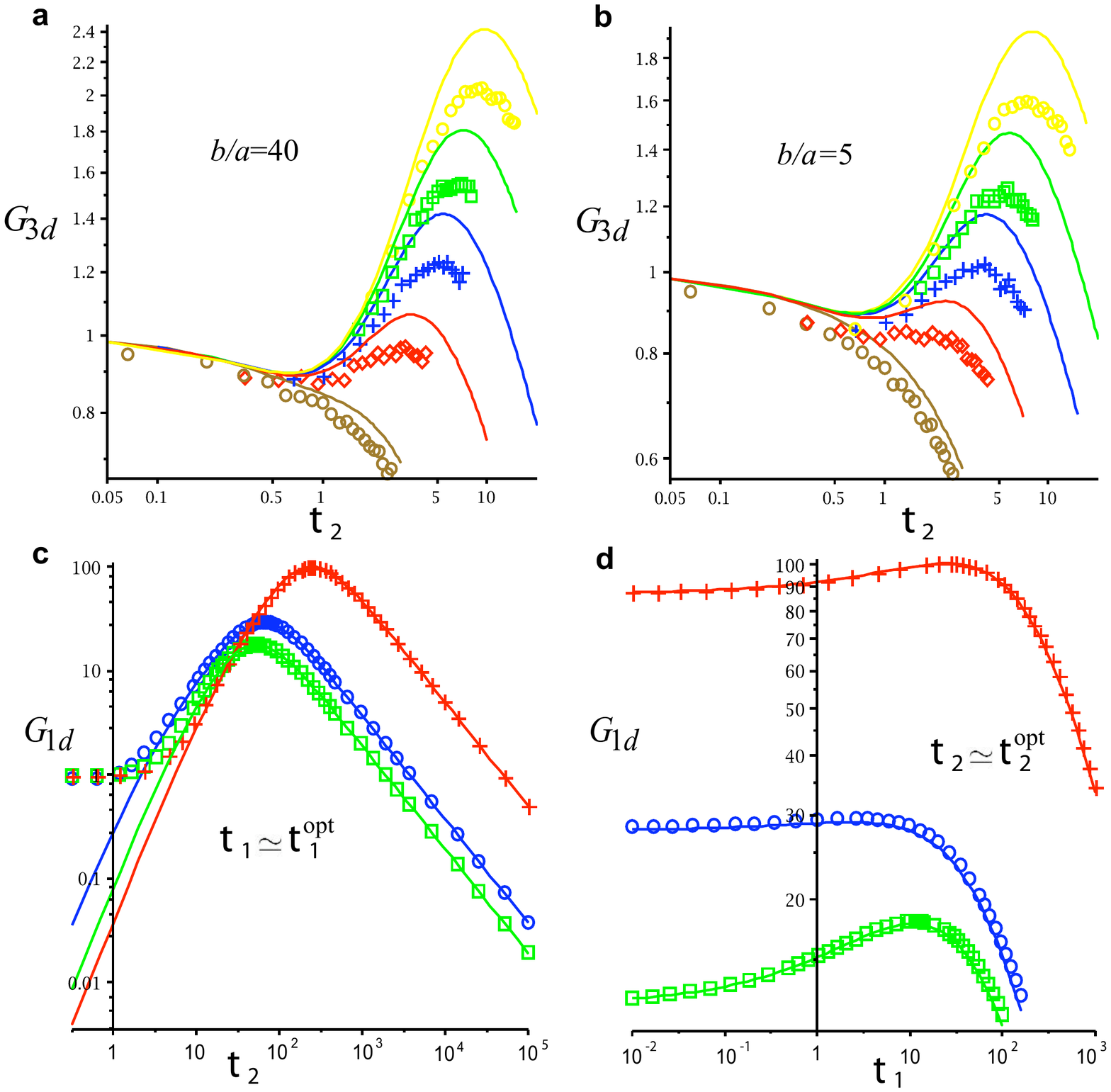}
\caption{}
\end{figure}

\end{document}